# Surface Oxidation of SnTe Analyzed by Self-Consistent Fitting of all Emission Peaks in its X-ray Photoelectron Spectrum

Martin Wortmann[1,2,‡,*], Beatrice Bednarz[3,‡], Negin Beryani Nezafat[4,5], Klaus Viertel[2], Olga Kuschel[3,6], Jan Schmalhorst[1], Inga Ennen[1], Maik Gärner[1], Natalie Frese[2], Gerhard Jakob[3], Joachim Wollschläger[6], Gabi Schierning[4,5,7], Andreas Hütten[1], Timo Kuschel[1,3]

[1] Bielefeld University, Faculty of Physics, Universitätsstraße 25, 33615 Bielefeld, Germany
[2] Bielefeld University of Applied Sciences and Arts, Faculty of Engineering and Mathematics, Interaktion 1, 33619 Bielefeld, Germany
[3] Johannes Gutenberg University Mainz, Institute of Physics, Staudingerweg 7, 55128 Mainz, Germany
[4] University of Duisburg-Essen, Institute for Energy and Materials Processes Applied Quantum Materials, 47057 Duisburg, Germany
[5] Research Center Future Energy Materials and Systems, Research Alliance Ruhr, 44780 Bochum, Germany
[6] Osnabrück University, Faculty of Physics, Barbarastraße 7, 49076 Osnabrück, Germany
[7] Center for Nanointegration Duisburg-Essen (CENIDE) and Nano Energie Technik Zentrum (NETZ), 47057 Duisburg, Germany

[‡] equal author contribution
[*] Correspondence: mwortmann@physik.uni-bielefeld.de

## Abstract

X-ray photoelectron spectroscopy (XPS) is among the most widely used methods for surface characterization. Currently, the analysis of XPS data is almost exclusively based on the main emission peak of a given element and the rest of the spectrum is discarded. This makes quantitative chemical state analyses by peak fitting prone to substantial error, especially in light of incomplete and flawed reference literature. However, most elements give rise to multiple emission peaks in the x-ray energy range, which are virtually never analyzed. For samples with an inhomogeneous depth distribution of chemical states, these peaks show different but interdependent ratios of signal components, as they correspond to different information depths. In this work, we show that self-consistent fitting of all emission peaks lends additional reliability to the depth profiling of chemical states by angular-resolved (AR-)XPS. We demonstrate this using a natively oxidized thin film of the topological crystalline insulator tin telluride (SnTe). This approach is not only complementary to existing depth profiling methods, but may also pave the way towards complete deconvolution of the XPS spectrum, facilitating a more comprehensive and holistic understanding of the surface chemistry of solids.

**Graphical Abstract**

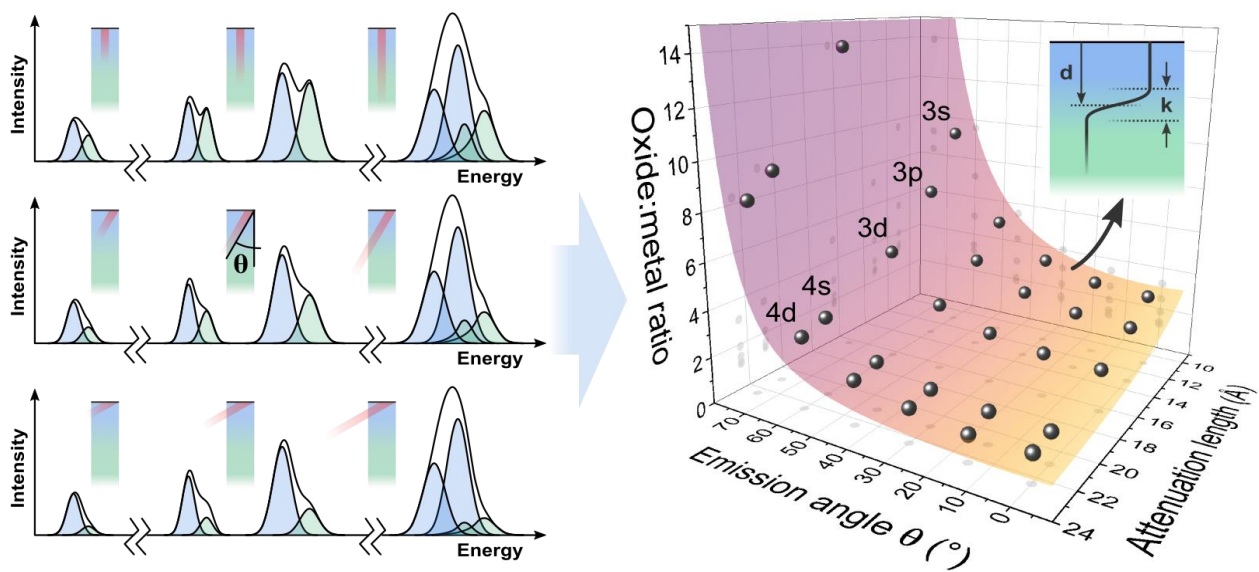

# 1. Introduction

X-ray photoelectron spectroscopy (XPS), originally called electron spectroscopy for chemical analysis (ESCA)[1], is today among the most widely used analytical methods in many areas of the natural sciences. More than 10,000 publications containing XPS data are published annually.[1][3] Since the inelastic mean free path $\lambda$ of electrons is in the range of only a few nanometers, it is primarily used to analyze surface chemistry, including elemental composition, chemical states, and within its information depth also depth distribution.[4][5]

A variety of methodologies have been utilized to derive depth information from XPS data. One such method involves the incremental surface ablation by ion etching. However, this method's depth resolution is limited, because of uncertainty of ablation rates, preferential sputtering, alterations of chemical states and intermixing perpendicular to the surface.[6][7][8] Non-destructive approaches to extract depth-information from XPS spectra are based purely on data modeling, leveraging relative signal intensities and known (i.e. calculable) depth distribution functions $\varphi(z)$, which represent the likelihood that an electron originates from a specific depth $z$ beneath the surface.[9] The nominal thickness $d$ of a single overlayer can be calculated from the measured intensity ratio $I_{overlayer}:I_{substrate}$ determined from the most intense emission peak(s) in a single fixed-angle spectrum (Figure 1).[10][11] This widely used method will be referred to as the single-energy-model, as it is usually relying on only a single intensity ratio with a specific $\varphi(z)$. When the layers consist of different elements, the most intense peak per element is used together with its relative sensitivity factor (RSF).[12] When the layers are from the same element in different chemical states, like in native oxide layers, peak fitting can be used to distinguish signal components within a peak region without the need for RSFs (Figure 1).[13]

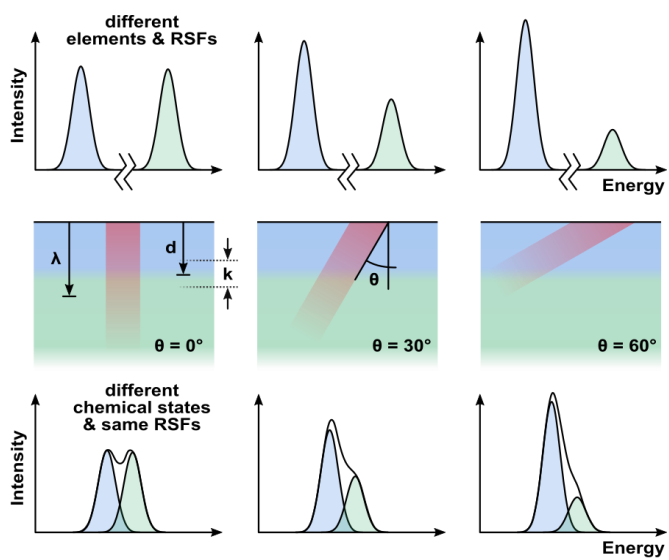

**Figure 1.** Schematic illustration of peak intensity ratios $I_{overlayer}:I_{substrate}$ at three different emission angles $\theta$ for a single overlayer (blue) with a thickness $d$ and interface sharpness $k$. The depth distribution function $\varphi(z)$ is illustrated as red bars. As the emission angle increases, surface sensitivity and overlayer-to-bulk signal ratio increases.

Angular-resolved XPS (ARXPS) can be used to obtain multiple intensity ratios from the same sample, since $\varphi(z)$ depends on the emission angle $\theta$ and the kinetic energy of the electrons $E_k$ (or rather $\lambda(E_k)$). Various models, such as the Multi-Layer-Method[14], Tikhonov regularization[15], or the Maximum-Entropy-Model[16], have been proposed to determine concentration profiles from ARXPS data. All these models rely only on the most intense emission peak(s) in the XPS spectrum (one per element), which are the only peaks for which peak fitting models are available in literature. For depth profiling of chemical states, the peak fitting is a significant contributor to the overall uncertainty, because the intensity ratio largely depends on the choice of background and line shapes (see Figure S1 in the supporting information (SI)), for which there is often no consensus.[17]

Each energy level in the x-ray energy range gives rise to at least one distinct emission peak in the spectrum. Albeit the lower-intensity peaks are virtually never analyzed. However, it has been shown that the consideration of an additional emission peak can improve the accuracy of concentration[18] and overlayer thickness measurements.[19] In an earlier publication[20], we have demonstrated that it is even possible to obtain a gradual concentration profile of oxidation states from a single fixed-angle spectrum when all emission peaks are utilized. In this work, we revise and apply this multiple-energies-model to ARXPS measurements of the binary alloy tin telluride (SnTe). We derive the oxide depth profile of a SnO overlayer on a SnTe thin film from self-consistent fitting of all emission peaks at multiple angles. This approach demonstrates enhanced reliability of the peak fitting and chemical state depth profiling. It is, in principle, complementary to other ARXPS depth profiling models and represents a possible route towards complete deconvolution of the XPS spectrum.

SnTe has recently been widely studied for its thermoelectric performance[21], its utility as topological insulator for spintronic and quantum computing applications[22], and its narrow bandgap for advanced optoelectronic devices[23]. From a practical standpoint, the presence and structure of native oxide layers and their effect on surface states and interface properties is rarely addressed. However, this work will focus exclusively on the proposed methodology.

## 2. Experimental

SnTe thin films of 50 nm nominal thickness were deposited onto Si wafers (with a 50 nm $SiO_2$ overlayer; cleaned with a stream of nitrogen) by radio frequency magnetron sputtering of a SnTe composite target (3" diameter, 99.99% purity) at room temperature with a source power of 15 W. At the time of measurement, the samples were exposed to standard lab atmosphere for 1 month in a polystyrene sample container.

The sample surface was imaged by scanning electron microscopy (SEM) using a LEO 1530 Gemini (Zeiss) operated at 20 kV. Transmission electron microscopy (TEM) was performed using a JEM 2200FS (JEOL) operated at 200 kV. The cross-section lamella has been prepared using a Helios dual beam FIB (FEI/Thermo Fisher) operated at 30 kV beam energy with a subsequent ion polishing step at 5 kV.

X-ray reflectometry (XRR) measurements were performed with a X'Pert Pro MPD PW3040-60 diffractometer (PANalytical) using Cu $K_\alpha$ radiation.

XPS was performed in a VersaProbe III device (PHI) at $3\times10^{-9}$ mbar using monochromatic Al $K_\alpha$ irradiation and a hemispherical electron analyzer (work function 4.39 eV) in constant analyzer energy mode at emission angles of 75°, 60°, 45°, 30°, 15°, and 0° relative to the surface normal. The angle between source and analyzer was 45° and the x-ray focus spot on the sample had a diameter of about 100 µm. The spectra were recorded with a pass energy of 55 eV (224 eV for surveys) and analyzed in CasaXPS Version 2.3.22. To compensate for charging effects and the unknown work function of the sample, the spectra were shifted so that the Fermi cutoff in the valence band coincides with a binding energy of 0 eV.[24]

Electron-excited Auger electron spectroscopy (AES) was performed in a PHI-660 Scanning Auger Microprobe (Perkin-Elmer). A depth profile was measured from 80 sequences of scanning with a 10 keV primary electron beam and etching by 500 eV $Ar^+$ ions for 90 s at 70° to the surface normal. The sample was rotated during ion etching to maximize depth resolution.[25]

## 3. Results

Figure 2a shows a schematic illustration of the surface of a SnTe thin film sample with a SnO overlayer. Throughout this document, the metallic SnTe will always be shown in green and the oxide SnO in blue. A top-view SEM image of the surface (Figure 2b) shows a granular topography indicating polycrystalline growth. Those grains of <50 nm are also seen in the TEM cross-section images at different levels of magnification (Figure 2c-e). The close-up TEM image (Figure 2c) shows the oxide layer on top and the polycrystalline SnTe underneath. The stoichiometry of the SnO layer has been confirmed to be 1:1 by XPS and AES, as will be discussed. It can be seen that the thickness of the oxide layer cannot be reliably determined from the TEM images. The main reason for this is that each TEM image necessarily represents an averaging over the thickness of the lamella sample, thus, making thickness measurements especially ambiguous for rough samples (this is illustrated in Figure S2). A rather unclear and subjective evaluation of the TEM images yields an average thickness of $d_{TEM} \approx 2.8$ nm of the SnO overlayer. Roughness is known to influence ARXPS depth profiling due to possible shadow casting at low emission angles.[26] The steepest deviation observed from an otherwise mostly flat surface measured 18° (see Figure 2d,e). For the sake of simplicity, a possible influence of roughness (i.e. by shadow casting at high emission angle) and lateral inhomogeneity (like grain boundaries) on depth profiling is considered negligible in the following.

For a comparison, XRR data has been analyzed, as shown in Figure 2f, which yields a thickness of the oxide overlayer of $d_{XRR} \approx 2.5$ nm. The layer stack underlying the model and the resulting scattering length density (SLD) plot is shown in the inset. It includes an intermediate SnO layer between the film and the substrate's SiO$_2$ surface, which has been confirmed by AES, too, as will be discussed. Although the presumption of the oxide layers significantly improves the accuracy of the XRR model, the exact values are highly unreliable, especially as SnO and SnTe have very similar nominal bulk (electron-)densities ($\rho_{t-SnO} \approx 6.29$ g/cm³ $\triangleq 1.63$ e⁻/Å³ and $\rho_{c-SnTe} \approx 6.32$ g/cm³ $\triangleq 1.59$ e⁻/Å³ according to materialsproject.org[27]). There are multiple parametrizations that are physically plausible and still mostly consistent with the data (the fit parameters for the model shown in Figure 2f are given in Table S1 in the SI).

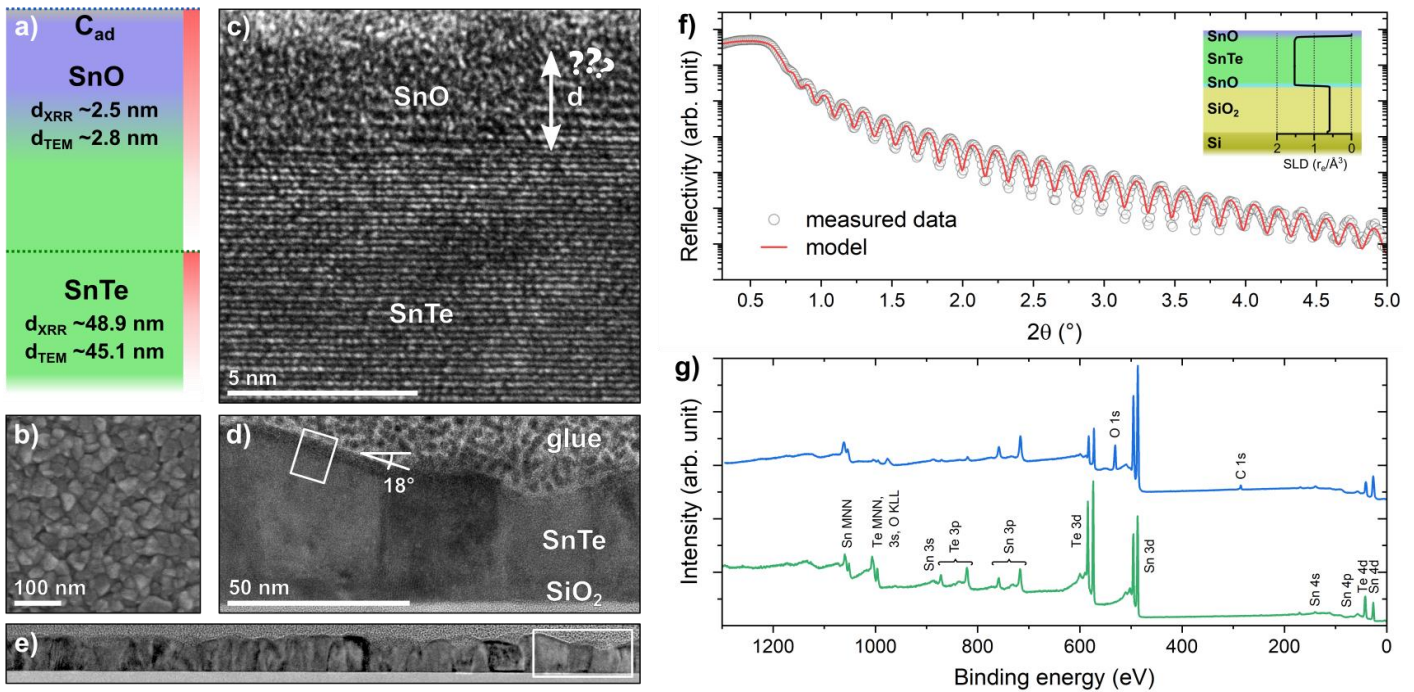

**Figure 2.** Characterization of the SnTe thin film sample: **(a)** Schematic illustration of the sample cross-section. The depth distribution functions of XPS signal intensity are illustrated as red bars for the pristine (blue dotted line) and sputtered (green dotted line) sample surface. Mind that the information depth is different for each emission peak and for each emission angle. **(b)** SEM image of the surface. **(c-e)** TEM cross-section images. The top image (c) is a magnification of the region indicated by the white rectangle in the image below (d), which is itself a magnification of the narrow image below that (e). **(f)** XRR data and model based on the layer stack illustrated in the inset. **(g)** XPS wide-scan spectra of the pristine oxidized surface (blue) and sputtered-etched reference measurement (green).

The XPS wide-scans in Figure 2g show multiple emission peaks from Sn and Te as well as O 1s and C 1s (adventitious carbon on the pristine surface). As a reference, one sample was sputtered, such that its surface contains no carbon and oxygen. This allows for an accurate peak assignment, as well as determination of binding energies and line shapes of the metallic component peaks. All peak models shown in Figure 3 except for Te 3d, Sn 3d, and O 1s are, to the best of our knowledge, novel and have so far not been thoroughly analyzed in existing literature.

In the pristine sample, the signal contribution of oxidized Sn $I_{Sn^{2+}}$ is very high compared to that of metallic Sn $I_{Sn^0}$, indicating a rather thick SnO layer. The nominal concentration of Te is less than 25 at% at $\theta = 0°$ and less than 10 at% at $\theta = 75°$. Its signal is completely metallic, indicating that there is no significant amount of Te in the oxide layer. The reference spectrum of the sample after sputtering had a Sn:Te atomic ratio of 1:1, as calculated from the Te 3d and Sn 3d emission peaks, within the uncertainty of the background subtraction.

The nominal oxide layer thickness $d$ can be calculated using the measured intensity ratio $I_{Sn^{2+}}/I_{Sn^0}$ of any given emission peak, typically the one with the highest intensity:

$$d = L_{SnO} \cdot \cos\theta \cdot \ln\left(\frac{N_{SnTe}}{N_{SnO}} \cdot \frac{L_{SnTe}}{L_{SnO}} \cdot \frac{I_{Sn^{2+}}}{I_{Sn^0}} + 1\right), \qquad 1$$

with $N_{SnO}$ and $N_{SnTe}$ being the atomic number densities of $Sn^{2+}$ in the oxide and $Sn^0$ in the metal, respectively. $L_{SnTe}$ and $L_{SnO}$ are known or calculated values of the effective attenuation length (EAL) in the oxide and in the metal. They are functions of the kinetic energy $E_k = h\nu - E_b - \phi$, for which $h\nu$ is the X-ray energy (here Al K$_\alpha$ with 1486.6 eV), $E_b$ is the binding energy, and $\phi$ is the work function of the spectrometer. It is widely recognized that the EAL is more accurate than the inelastic mean free path (IMFP) for quantitative XPS depth profiling, as it takes into account the influence of elastic scattering.[28][29] Therefore, in the following, only $L$ is used instead of $\lambda$. Additional discussion and information on the calculation of both $L$ and $\lambda$ is given in the SI. In principle, eq. 1 should give the same result for every peak and emission angle. Different intensity ratios are therefore expected, because different peaks correspond to different values of $E_k$.

In an earlier publication, we advanced this widely used single-energy model for the application to multiple emission peaks in a fixed-angle spectrum (each consisting of one metallic and one oxide component peak) to obtain a gradual concentration profile at the oxide-metal interface, as described by an error function $C_{Sn^{2+}}(z) = 0.5 \cdot \text{erfc}(k \cdot (z - d)) \cdot 100\%$, for which $k$ is the sharpness of the interface:[20]

$$\frac{I_{Sn^{2+}} \cdot N_{SnTe}}{I_{Sn^0} \cdot N_{SnO}}(L_{av}, \theta) = \frac{\text{erfc}(-kd) - \text{erfc}\left(\frac{1}{2L_{av}k \cdot \cos\theta} - kd\right) \cdot \exp\left(\frac{1}{(2L_{av}k \cdot \cos\theta)^2} - \frac{d}{L_{av} \cdot \cos\theta}\right)}{\text{erfc}(kd) + \text{erfc}\left(\frac{1}{2L_{av}k \cdot \cos\theta} - kd\right) \cdot \exp\left(\frac{1}{(2L_{av}k \cdot \cos\theta)^2} - \frac{d}{L_{av} \cdot \cos\theta}\right)}. \qquad 2$$

In this equation, $L_{av}$ is the average of $L_{SnTe}$ and $L_{SnO}$. However, this averaging can be subject to error if $L_{SnTe}$ and $L_{SnO}$ differ sufficiently from each other. We have therefore revised the equation[20] to include distinct values for the metal and the oxide. The resulting model has three independent variables $L_{SnO}$, $L_{SnTe}$, and $\theta$ as well as two fitting parameters $k$ and $d$ (additional information on the derivation is given in the SI):

$$\frac{I_{Sn^{2+}} \cdot N_{SnTe}}{I_{Sn^0} \cdot N_{SnO}}(L_{SnO}, L_{SnTe}, \theta)$$

$$= \frac{L_{SnO}}{L_{SnTe}} \cdot \exp\left(\frac{d}{L_{SnO} \cdot \cos\theta} - \frac{d}{L_{SnTe} \cdot \cos\theta}\right)$$

$$\cdot \frac{\text{erfc}(-kd) - \text{erfc}\left(\frac{1}{2L_{SnO}k \cdot \cos\theta} - kd\right) \cdot \exp\left(\frac{1}{(2L_{SnO}k \cdot \cos\theta)^2} - \frac{d}{L_{SnO} \cdot \cos\theta}\right)}{\text{erfc}(kd) + \text{erfc}\left(\frac{1}{2L_{SnTe}k \cdot \cos\theta} - kd\right) \cdot \exp\left(\frac{1}{(2L_{SnTe}k \cdot \cos\theta)^2} - \frac{d}{L_{SnTe} \cdot \cos\theta}\right)}. \qquad 3$$

This equation reduces to eq. 1 in the limit $k \rightarrow \infty$, i.e. a step function profile at the interface. Both eq. 2 and 3 can either be applied to a single spectrum at constant $\theta$ or to an ARXPS data set for which $\theta$ is treated as an independent variable. Here, we utilize both variants at once, thus mapping the entire energy-angle landscape. Peak fitting models were developed for all emission peaks and applied to spectra measured at emission angles of $\theta = 75°, 60°, 45°, 30°, 15°, 0°$. Figure 3 shows most of these peak fittings at three exemplary angles (some peaks and some angles have been omitted for better readability). A more detailed discussion on the development of these peak models as well as binding energy values, line shapes, and backgrounds can be found in the SI. The peak fittings yield the $I_{Sn^{2+}}/I_{Sn^0}$ intensity ratios for every emission peak of Sn. Te shows no sign of oxidation.

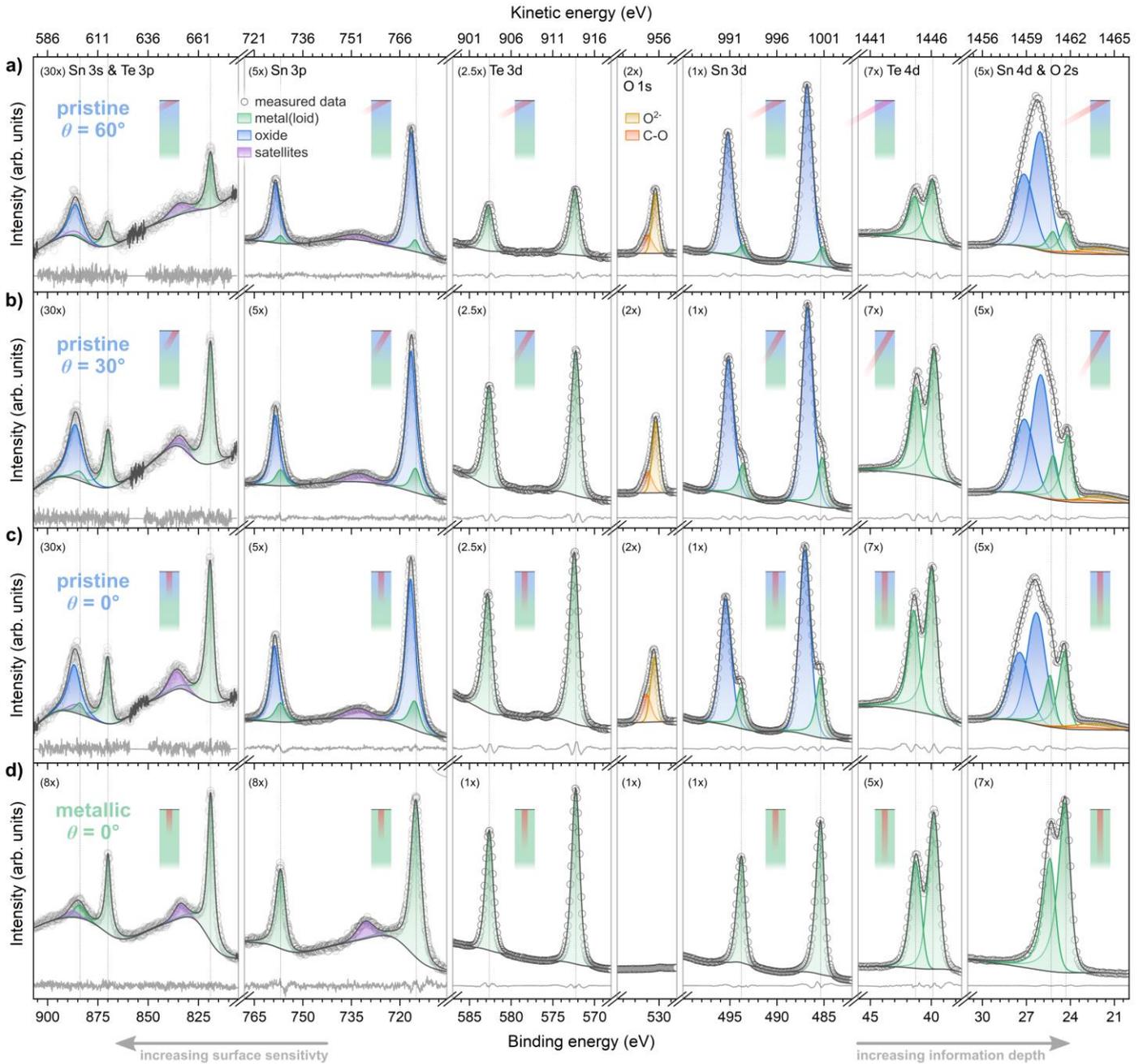

**Figure 3.** XPS narrow-scans of all emission peaks, except for Sn 4s, Te 4s, C 1s, Auger peaks, and valence band (Sn 4p is unresolvable). These peak regions as well as the intermediate emission angles $\theta = 75°, 45°, 15°$ have been omitted in this figure for better readability. The depth distribution function for each emission peak is illustrated schematically in the insets (red bars). For better visibility, the peaks have been magnified by a factor indicated in brackets in the top left-hand corner of each peak region. The fit residuals are shown in pale grey below the spectra.

These intensity ratios are then weighted by atomic number densities $(I_{Sn^{2+}} \cdot N_{SnTe})/(I_{Sn^0} \cdot N_{SnO})$. Each peak occurs at a specific value of $E_k$ corresponding to $L_{SnTe}$ and $L_{SnO}$ (for spin orbit doublets the weighted average of the peak position was used as $E_k$). The weighted intensity ratios have then been fitted by eq. 3 to determine the parameters $d$ and $k$, which is shown as a set of colored surface plots in Figure 4a. However, not all peak fittings are equally reliable. The peaks with the highest intensity and clearest chemical shift between $Sn^{2+}$ and $Sn^0$ are 3d and 4d. The *s*- and *p*-orbitals show no significant chemical shift (in contrast to other elements such as Ti, for example, for which even the *s*-orbitals show a significant chemical shift[20]). The fitting of eq. 3 is thus only based on 3d and 4d. The intensity ratios of the other emission peaks have then been taken from the parameterized fit function and applied as constraints in the XPS peak fitting (after unweighting $N_{SnTe}/N_{SnO}$). In this way, these peaks do not contribute any significant information to the depth profiling, but conversely, their fitting is more stringently constrained by the expected intensity ratios. Thus, reliable and self-consistent peak fitting models can be derived from the depth profiling. The same intensity ratios were then plotted as a function of $\theta$ and $L_{av}$ and fitted by eq. 2 to test the influence of averaging. The resulting values of $d$ and $k$ are given in Figure 4 and illustrated in Figure 5c,d. It should be mentioned that EAL values for $\theta > 60°$ are increasingly error-prone and should not be applied on their own.[30] All unweighted intensity ratios are given in Table S3 in the SI.

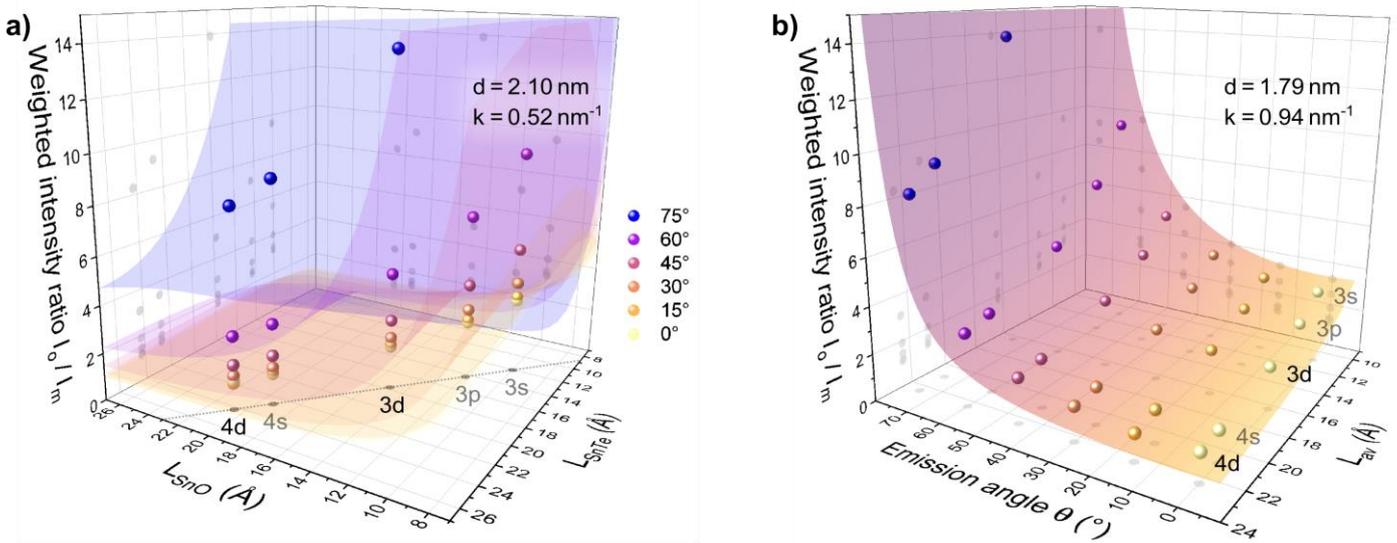

**Figure 4. (a)** Intensity ratios weighted by atomic number densities $(I_{Sn^{2+}} \cdot N_{SnTe})/(I_{Sn^0} \cdot N_{SnO})$ as a function of the emission angle $\theta$, $L_{SnO}$, and $L_{SnTe}$. The set of colored surface plots is a 3D-representation of the parameterized eq. 3. **(b)** The same intensity ratios as in (a) as a function of $\theta$ and $L_{av}$, according to eq. 2. The parameters $d$ and $k$ correspond to the average SnO film thickness and the sharpness of the interface, respectively. The data points from 3p and 3s at $\theta = 75°$ are not shown to enhance the clarity of the diagrams.

Figure 5a shows the depth profiles of $Sn^{2+}$ and $Sn^0$ obtained from eq. 3. The nominal oxide layer thickness is 2.10 nm, however, the gradual transition region spans more than twice as deep. To confirm the validity of these results, Figure 5b and c show the depth profiles obtained from a sequence of 80 iterations of AES and 90 s of $Ar^+$ sputtering. The first 20 spectra that contain information about the native oxide layer are shown in Figure 5b. While there is a clear chemical shift of the Sn MNN peak region, the Te MNN peak changes only in intensity apart from the vanishing overlap with the O $KL_1L_1$ and $KL_1L_{2-3}$ peaks. The main O $KL_{2-3}L_{2-3}$ peak vanishes as the oxide layer is ablated. In the first spectrum at the surface, there is adventitious carbon with some carbon-bonded oxygen (C KLL not shown). Below the contamination, the results indicate a stoichiometric 1:1 SnO layer – in contrast to $SnO_2$, which is a stable Sn oxide, too. This agrees with the XPS data, which shows an atomic ratio of $Sn^{2+}$ ions in Sn $3d_{5/2}$ to $O^{2-}$ in O 1s of approximately 1:1. Since, in this case, the XPS models do not distinguish between different oxidation states and the signals are accurately fitted by a single Voigt function, a possible contribution of $SnO_2$ is not considered here. In addition, the AES spectra show that there is indeed no Te in the surface of the oxide layer. The signal in the Te MNN peak region is fully explained by the less intense O KLL transitions, taking into account their known relative intensity ratios.

Figure 5c visualizes these changes in the relative elemental composition as a function of sputtering duration according to the quantitative evaluation of the AES spectra. Mind that it is not possible to accurately determine the ablation rate in layers with varying composition, which is why the sputter time was not converted to a thickness. The apparent composition might also be influenced by preferential sputtering. Most importantly, each data point is averaged over its information depth, which is about 3 nm (i.e. $\lambda \approx 1$ nm) for the Auger electrons in both SnO and SnTe. Possible differences between the sensitivity factors for the metal and oxide are also neglected. For quantification in AES the raw spectrum is differentiated and the peak-to-peak height – measured as the difference between the maximum and minimum around the Auger peak – is used as the intensity, which is then normalized with RSFs to obtain relative atomic concentrations. The AES oxygen depth profile nevertheless shows very good agreement with the XPS results presented in Figure 5a.

Underneath the oxide layer, there is a region of reduced Sn concentration, from which the excess Sn in the oxide layer presumably originates. Again, preferential sputtering might contribute to this apparent discrepancy. The nominal 1:1

SnTe ratio is only reached after 70 min of sputtering. The measurement also confirms the presence of a SnO layer between the film and the substrate, indicated by a grey arrow.

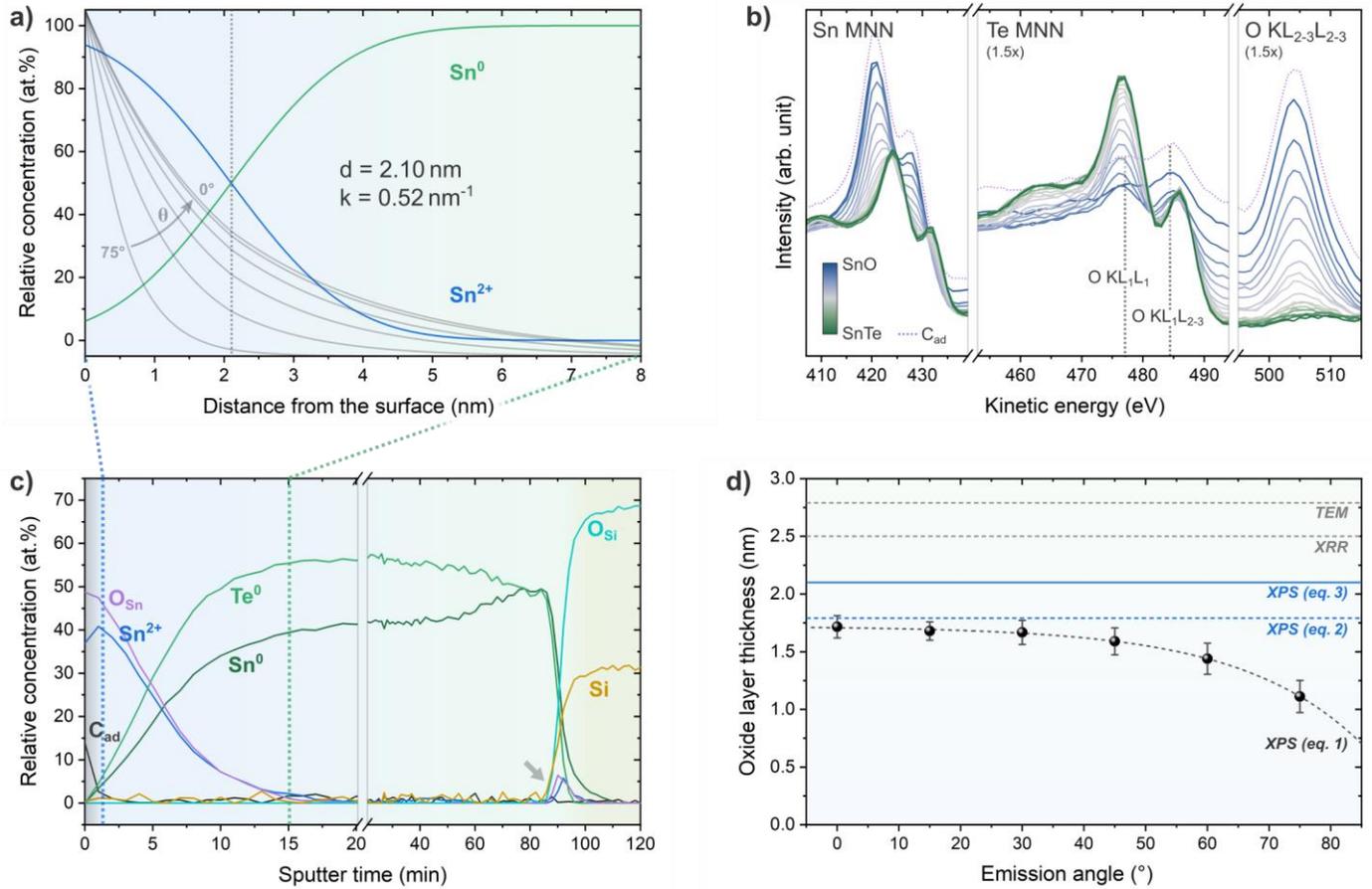

**Figure 5. (a)** Relative proportion of $Sn^{2+}$ and $Sn^0$ as a function of the distance from the surface, i.e. the oxide depth profile resulting from the fit shown in Figure 4a. The grey plots are the depth distribution functions $\varphi(z)$ of Sn 3d for different emission angles with a discontinuity at $z = d$. **(b)** The first 20 spectra from a sequence of 80 iterations of AES and subsequent $Ar^+$ sputtering (1st spectrum: dotted line; 2nd spectrum: dark blue; 20th spectrum: dark green). A comparison of the Auger emission peaks recorded via AES and ARXPS is shown in Figure S5 in the SI. **(c)** Elemental depth profile as a function of sputter time. The x-axis scaling is enlarged on the left side. The SnTe thin film had a nominal thickness of 50 nm. The SnTe layer without oxide layers is $d_{XRR} \approx 46\ nm$ corresponding to a sputtering rate in SnTe of about $1.7\ nm/min$. **(d)** Average oxide overlayer thickness according to the single-energy-model (eq. 1) as a function of the emission angle in comparison with the results of XRR, TEM, and the multiple-energies model with averaged (eq. 2) and distinct (eq. 3) EALs. The error bars are the standard deviation of the results from different emission peaks in the same spectrum (calculated with the intensity ratios seen in Figure 4).

Figure 5d summarizes the results obtained for the average oxide layer thickness obtained from the different methods. The nominal oxide layer thickness determined by XPS yielded lower results as XRR and TEM, however, in this specific case, both methods turned out to be unreliable and sensitive to subjective analysis procedure. It should be reiterated that the results of XRR and TEM shown in Figure 5d are consistent with the data over a wide range of values. The single-energy model (eq. 1) gave angle-dependent oxide layer thicknesses for the same intensity ratios as used in eq. 3. The thickness values appear to decrease from 1.7 nm at low emission angle to 1.1 nm at high emission angle, which is lower than the result of eq. 3 for all angles. The difference between the two methods is their assumption about the interface sharpness. If, contrary to the single-energy model assumption, the oxide layer has a broad interface, only alloys of unknown stoichiometry instead of distinct layers lie within the information depth (as illustrated by the overlaid depth distribution functions in Figure 5c). Consequently, eq. 1 systematically underestimates the layer thickness, with the deviation increasing at higher emission angles and with broader concentration gradients at the interface (i.e. decreasing $k$). This is demonstrated in Figure S6 in the SI. The results indicate that eq. 1 is only reliable for sharp interface profiles with $k > 1$. Another contributing factor may be that the EAL increases at large angles due to elastic scattering. Usually, EAL values are considered unreliable for $\theta > 60°$.[3]

Although the use of multiple energies and angles significantly reduces the ambiguity, we still estimate the range of thickness values that are reasonably compatible with the data to be about ±0.25 nm, which is, in the present case, significantly less than for TEM and XRR.

## 4. Discussion

The presented results highlight the high reliability of oxygen depth profiles obtained by XPS analysis using self-consistent fitting of all emission peaks at multiple angles. They also emphasize the significant errors associated with the single-energy-model when analyzing samples with broad concentration gradients. Therefore, it is very important to be aware of the challenges involved in XPS data analysis.

Peak fitting is subject to a variety of physical and empirical constraints, some of which are physically determined (like the intensity ratio of spin-orbit split doublets), sample- or device-dependent (like peak position or energy resolution), widely debated (basically every detail), or completely unknown (like every emission peak besides the most intense in the spectrum).[31][32] Achieving a consistent and physically meaningful fit is a demanding and iterative process that often does not lead to unambiguous and unique solutions, even when done by experts with the aid of published reference spectra. The use of all emission peaks instead of just one has three main advantages: 1. The reliability of peak fitting is significantly improved because all peak models within a spectrum must be self-consistent. 2. The need for self-consistency allows previously unexplored spectral regions (i.e. the entire spectrum with the exception of the main emission peak) to be elucidated with high confidence. 3. Depth profiling of chemical states is more reliable as it is based on two independent variables ($L$ and $\theta$) instead of just one ($\theta$).

With regard to 1. and 2.: The increasing accessibility of XPS, as reflected by the rapid increase in the number of publications[1], has led to widespread misuse, incorrect data analysis, and insufficient data reporting, with about 40% of published peak fitting models containing major errors.[33][34][35][36] The widespread proliferation of flawed data and data analysis compromises the integrity of the scientific literature, misguiding researchers and perpetuating incorrect methodologies across studies.[37] These problems relate almost exclusively to the most dominant emission peak of a given element, because peaks with lower intensity are almost never considered. For those peaks, there are usually not even reference spectra available in data repositories such as xpsfitting.com, xpsoasis.org, xpsdatabase.net, or xpslibrary.com. Peak assignment based on binding energy values from commercial software or published literature is highly unreliable, because for decades demonstrably incorrect methods were used to reference the binding energy scale.[38] On the one hand, the lack of supporting scientific literature is a challenge for the multiple-energies approach. On the other hand, the interdependency of intensity ratios represents a significant opportunity as it facilitates the unambiguous identification and modeling of component peaks.

The multiple-energies approach is in principle complementary to existing ARXPS depth profiling models as it just expands the amount of underlying information. The depth profile model presented here is conceptually related to the Multi-Layer-Method[14] that also parameterizes the depth profile of each species as a layer with defined stoichiometry. An essential difference is that, in this work, the interface between two layers is not assumed to be abrupt but gradual, described by a sigmoid function (like the complementary error function) with an additional parameter $k$. However, the model in its current form was only applicable to the specific case of SnTe/SnO, because the $Sn^{2+}/Sn^0$ ratio adequately describes the entire depth profile. Further research is required to extend the multiple-energies approach to multi-layer stacks of different materials. Apart from depth profiling, a better understanding of less pronounced emission peaks also allows more precise evaluation of composite materials in which the peaks of different elements may overlap.

Assuming that the fundamental sequence of layers and their stoichiometry are known, which is often provided either by the production process or by other measurement methods, it is in principle possible to reconstruct the entire XPS spectrum. All emission peaks are intrinsically interdependent and the signal background is also contingent on the structure of the sample.[39] Models that incorporate this intrinsic interdependence represent a first step towards complete spectral deconvolution, which may ultimately facilitate complete knowledge of the surface chemistry. The widespread failure to acknowledge internal dependencies and constraints within the XPS spectrum represents a largely unexplored field with significant potential for analytical surface science and implications for many domains of applied research. It is quite conceivable that software with reasonably initialized parameters – possibly by means of artificial intelligence[40] – will someday automate self-consistent reconstruction of the entire spectrum by combining background analysis with a multiple-energies approach, which may alleviate the legacy of decades of erroneous scientific literature.

## 5. Conclusion

We have introduced a new depth profiling model for ARXPS based on the self-consistent fitting of all emission peaks. This approach increases the reliability of chemical state depth profiling on the one hand and enables the elucidation of previously unexplored peak regions by exploiting internal dependencies on the other. The method was demonstrated using the example of a natively oxidized thin film of the topological crystalline insulator SnTe. Modeling of the ARXPS data resulted in a nominal thickness of the SnO oxide layer of about 2 nm with a broad interface spanning over more than 4 nm. The consideration of the entire spectrum enables a richer utilization of available data and a more holistic understanding of the surface chemistry, possibly even paving the way towards the complete deconvolution of the XPS spectrum.

**Acknowledgement**

The authors would like to thank Prof. Dr. Günter Reiss from Bielefeld University for providing access to his laboratory. This project has received funding from the European Union's Horizon 2020 Research and Innovation Programme under the Marie Skłodowska-Curie grant agreement No 860060 "Magnetism and the effect of Electric Field" (MagnEFi). G.S. gratefully acknowledges funding from the European Research Council (ERC) under the European Union's Horizon 2020 research and innovation program grant agreement No. 863823 MATTER. Further, parts of this project have received funding by the Deutsche Forschungsgemeinschaft (DFG, German Research Foundation) in Project-ID SCHI 1010/12-1.

**Declaration of competing interest**

The authors declare that they have no known competing financial interests or personal relationships that could have appeared to influence the work reported in this paper.

**Author contributions**

B.B., K.V., and M.W. derived and implemented the models; I.E., J.S., M.W., M.G., N.F., and O.K. performed measurements; N.B.N. prepared the samples; B.B. and M.W. did the data analysis; A.H., G.J., G.S., J.W., M.W., and T.K. planned and supervised the experimental work; M.W. wrote the manuscript with the assistance of all co-authors; B.B., A.H., G.J., G.S., J.W., M.W., and T.K. revised the manuscript.

**Data availability statement**

The data that support the findings of this study are available from the corresponding author upon reasonable request.

# References

[1] Pinder, J. W., Kulbacki, B., Baer, D. R., Biesinger, M., Castle, J., Castner, D. G., ... & Linford, M. R. (2025). What's in a Name?"ESCA" or "XPS"? A Discussion of Comments Made by Kai Siegbahn More Than Four Decades Ago Regarding the Name of the Technique. Surface and Interface Analysis.

[2] Greczynski, G., & Hultman, L. (2020). X-ray photoelectron spectroscopy: towards reliable binding energy referencing. Progress in Materials Science, 107, 100591.

[3] Jablonski, A., & Powell, C. J. (2020). Effective attenuation lengths for different quantitative applications of X-ray photoelectron spectroscopy. Journal of Physical and Chemical Reference Data, 49(3).

[4] Wortmann, M., Samanta, T., Gaerner, M., Westphal, M., Fiedler, J., Ennen, I., ... & Ehrmann, A. (2023). Isotropic exchange-bias in twinned epitaxial Co/Co3O4 bilayer. APL Materials, 11(12).

[5] Watts, J. F., Wolstenholme, J. (2019). An introduction to surface analysis by XPS and AES. John Wiley & Sons.

[6] Hofmann, S. (2014). Sputter depth profiling: past, present, and future. Surface and Interface Analysis, 46(10-11), 654-662.

[7] Hofmann, S., Zhou, G., Kovac, J., Drev, S., Lian, S. Y., Lin, B., ... & Wang, J. Y. (2019). Preferential sputtering effects in depth profiling of multilayers with SIMS, XPS and AES. Applied Surface Science, 483, 140-155.

[8] Biedinger, J., Yao, Z., Wortmann, M., Westphal, M., Frese, N., Gehra, R., Brandt, N., … & Reiss, G. (2024) $Al_2O_3$-Functionalized Carbon Nanomembranes with Enhanced Water Permeance and Selectivity for Efficient Air Dehumidification. Advanced Functional Materials. 2418806

[9] Powell, C. J. (2020). Practical guide for inelastic mean free paths, effective attenuation lengths, mean escape depths, and information depths in x-ray photoelectron spectroscopy. Journal of Vacuum Science & Technology A, 38(2).

[10] Krishna, D. N. G., & Philip, J. (2022). Review on surface-characterization applications of X-ray photoelectron spectroscopy (XPS): Recent developments and challenges. Applied Surface Science Advances, 12, 100332.

[11] Fadley, C. S., Baird, R. J., Siekhaus, W., Novakov, T., & Bergström, S. Å. L. (1974). Surface analysis and angular distributions in x-ray photoelectron spectroscopy. Journal of Electron Spectroscopy and Related Phenomena, 4(2), 93-137.

[12] Fadley, C. S. (2010). X-ray photoelectron spectroscopy: Progress and perspectives. Journal of Electron Spectroscopy and Related Phenomena, 178, 2-32.

[13] Cumpson, P. J. (1995). Angle-resolved XPS and AES: depth-resolution limits and a general comparison of properties of depth-profile reconstruction methods. Journal of Electron Spectroscopy and Related Phenomena, 73(1), 25-52.

[14] Herrera-Gomez, A., Guzman-Bucio, D. M., Mayorga-Garay, M., & Cortazar-Martinez, O. (2023). Angle resolved x-ray photoelectron spectroscopy assessment of the structure and composition of nanofilms—including uncertainties—through the multilayer model. Journal of Vacuum Science & Technology A, 41(6).

[15] Paynter, R. W. (2012). Regularization methods for the extraction of depth profiles from simulated ARXPS data derived from overlayer/substrate models. Journal of Electron Spectroscopy and Related Phenomena, 184(11-12), 569-582.

[16] Smith, G. C., & Livesey, A. K. (1992). Maximum entropy: A new approach to non-destructive deconvolution of depth profiles from angle-dependent XPS. Surface and Interface Analysis, 19(1-12), 175-180.

[17] Engelhard, M. H., Baer, D. R., Herrera-Gomez, A., & Sherwood, P. (2020). Introductory guide to backgrounds in XPS spectra and their impact on determining peak intensities. Journal of Vacuum Science & Technology A, 38(6).

[18] Jablonski, A., Lesiak, B., Zommer, L., Ebel, M. F., Ebel, H., Fukuda, Y., ... & Tougaard, S. (1994). Quantitative analysis by XPS using the multiline approach. Surface and interface analysis, 21(10), 724-730.

[19] Jablonski, A., & Zemek, J. (2009). Overlayer thickness determination by XPS using the multiline approach. Surface and Interface Analysis: An International Journal devoted to the development and application of techniques for the analysis of surfaces, interfaces and thin films, 41(3), 193-204.

[20] Wortmann, M., Viertel, K., Westphal, M., Graulich, D., Yang, Y., Gärner, M., ... & Kuschel, T. (2023). Sub-Nanometer Depth Profiling of Native Metal Oxide Layers Within Single Fixed-Angle X-Ray Photoelectron Spectra. Small Methods, 2300944.

[21] Moshwan, R., Yang, L., Zou, J., & Chen, Z. G. (2017). Eco-friendly SnTe thermoelectric materials: progress and future challenges. Advanced Functional Materials, 27(43), 1703278.

[22] Tanaka, Y., Ren, Z., Sato, T., Nakayama, K., Souma, S., Takahashi, T., ... & Ando, Y. (2012). Experimental realization of a topological crystalline insulator in SnTe. Nature Physics, 8(11), 800-803.

[23] Kovalenko, M. V., Heiss, W., Shevchenko, E. V., Lee, J. S., Schwinghammer, H., Alivisatos, A. P., & Talapin, D. V. (2007). SnTe nanocrystals: a new example of narrow-gap semiconductor quantum dots. Journal of the American Chemical Society, 129(37), 11354-11355.

[24] Greczynski, G., & Hultman, L. (2024). Binding energy referencing in X-ray photoelectron spectroscopy. Nature Reviews Materials, 1-17.

[25] Zalar, A. (1990). Significance of sample rotation in auger electronspectroscopy sputter depth profiling of thin films. Thin solid films, 193, 258-269.

[26] Martín-Concepción, A. I., Yubero, F., Espinos, J. P., & Tougaard, S. (2004). Surface roughness and island formation effects in ARXPS quantification. Surface and Interface Analysis: An International Journal devoted to the development and application of techniques for the analysis of surfaces, interfaces and thin films, 36(8), 788-792.

[27] Jain, A., Ong, S. P., Hautier, G., Chen, W., Richards, W. D., Dacek, S., ... & Persson, K. A. (2013). Commentary: The Materials Project: A materials genome approach to accelerating materials innovation. APL materials, 1(1).

[28] Powell, C. J., & Jablonski, A. (1999). Evaluation of calculated and measured electron inelastic mean free paths near solid surfaces. Journal of Physical and Chemical Reference Data, 28(1), 19-62.

[29] Cumpson, P. J., & Seah, M. P. (1997). Elastic scattering corrections in AES and XPS. II. Estimating attenuation lengths and conditions required for their valid use in overlayer/substrate experiments. Surface and Interface Analysis: An International Journal devoted to the development and application of techniques for the analysis of surfaces, interfaces and thin films, 25(6), 430-446.

[30] Powell, C. J., & Jablonski, A. (2002). Electron effective attenuation lengths for applications in Auger electron spectroscopy and x-ray photoelectron spectroscopy. Surface and Interface Analysis: An International Journal devoted to the development and application of techniques for the analysis of surfaces, interfaces and thin films, 33(3), 211-229.

[31] Major, G. H., Fernandez, V., Fairley, N., Smith, E. F., & Linford, M. R. (2022). Guide to XPS data analysis: Applying appropriate constraints to synthetic peaks in XPS peak fitting. Journal of Vacuum Science & Technology A, 40(6).

[32] Shard, A. G. (2020). Practical guides for x-ray photoelectron spectroscopy: Quantitative XPS. Journal of Vacuum Science & Technology A, 38(4).

[33] Linford, M. R., Smentkowski, V. S., Grant, J. T., Brundle, C. R., Sherwood, P. M., Biesinger, M.C., ... Baer, D. R. (2020). Proliferation of faulty materials data analysis in the literature. Microscopy and Microanalysis, 26(1), 1-2.

[34] Major, G. H., Pinder, J. W., Austin, D. E., Baer, D. R., Castle, S. L., Čechal, J., ... & Linford, M. R. (2023). Perspective on improving the quality of surface and material data analysis in the scientific literature with a focus on x-ray photoelectron spectroscopy (XPS). Journal of Vacuum Science & Technology A, 41(3).

[35] Major, G. H., Avval, T. G., Moeini, B., Pinto, G., Shah, D., Jain, V., ... & Linford, M. R. (2020). Assessment of the frequency and nature of erroneous x-ray photoelectron spectroscopy analyses in the scientific literature. Journal of Vacuum Science & Technology A, 38(6).

[36] Major, G. H., Clark, B. M., Cayabyab, K., Engel, N., Easton, C. D., Čechal, J., ... & Linford, M. R. (2023). Insufficient reporting of x-ray photoelectron spectroscopy instrumental and peak fitting parameters (metadata) in the scientific literature. Journal of Vacuum Science & Technology A, 41(4).

[37] Greczynski, G., & Hultman, L. (2017). C 1s peak of adventitious carbon aligns to the vacuum level: dire consequences for material's bonding assignment by photoelectron spectroscopy. ChemPhysChem, 18(12), 1507-1512.

[38] Greczynski, G., & Hultman, L. (2020). Compromising science by ignorant instrument calibration—need to revisit half a century of published XPS data. Angewandte Chemie, 132(13), 5034-5038. https://doi.org/10.1002/ange.201916000

[39] Tougaard, S. (2010). Energy loss in XPS: Fundamental processes and applications for quantification, non-destructive depth profiling and 3D imaging. Journal of Electron Spectroscopy and Related Phenomena, 178, 128-153.

[40] Pielsticker, L., Nicholls, R. L., DeBeer, S., & Greiner, M. (2023). Convolutional neural network framework for the automated analysis of transition metal X-ray photoelectron spectra. Analytica Chimica Acta, 1271, 341433.

# Supporting Information

# Analyzing Surface Oxidation of SnTe by Self-Consistent Fitting of all Emission Peaks in its X-ray Photoelectron Spectrum

Martin Wortmann[1,2,‡,*], Beatrice Bednarz[3,‡], Negin Beryani Nezafat[4,5], Klaus Viertel[2], Olga Kuschel[3,6], Jan Schmalhorst[1], Inga Ennen[1], Maik Gärner[1], Natalie Frese[2], Gerhard Jakob[3], Joachim Wollschläger[6], Gabi Schierning[4,5,7], Andreas Hütten[1], Timo Kuschel[1,3]

[1] Bielefeld University, Faculty of Physics, Universitätsstraße 25, 33615 Bielefeld, Germany
[2] Bielefeld University of Applied Sciences and Arts, Faculty of Engineering and Mathematics, Interaktion 1, 33619 Bielefeld, Germany
[3] Johannes Gutenberg University Mainz, Institute of Physics, Staudingerweg 7, 55128 Mainz, Germany
[4] University of Duisburg-Essen, Institute for Energy and Materials Processes Applied Quantum Materials, 47057 Duisburg, Germany
[5] Research Center Future Energy Materials and Systems, Research Alliance Ruhr, 44780 Bochum, Germany
[6] Osnabrück University, Faculty of Physics, Barbarastraße 7, 49076 Osnabrück, Germany
[7] Center for Nanointegration Duisburg-Essen (CENIDE) and Nano Energie Technik Zentrum (NETZ), 47057 Duisburg, Germany

[‡] equal author contribution
[*] Correspondence: mwortmann@physik.uni-bielefeld.de.de

## 1. Reliability of Peak Fitting

Figure S1 shows different peak fitting models of the Sn 3d peak region, which are about equally consistent with the measured XPS data, but result in very different intensity ratios. The example illustrates that relying on a single emission peak for the calculation of quantitative depth information introduces large uncertainty to the data analysis.

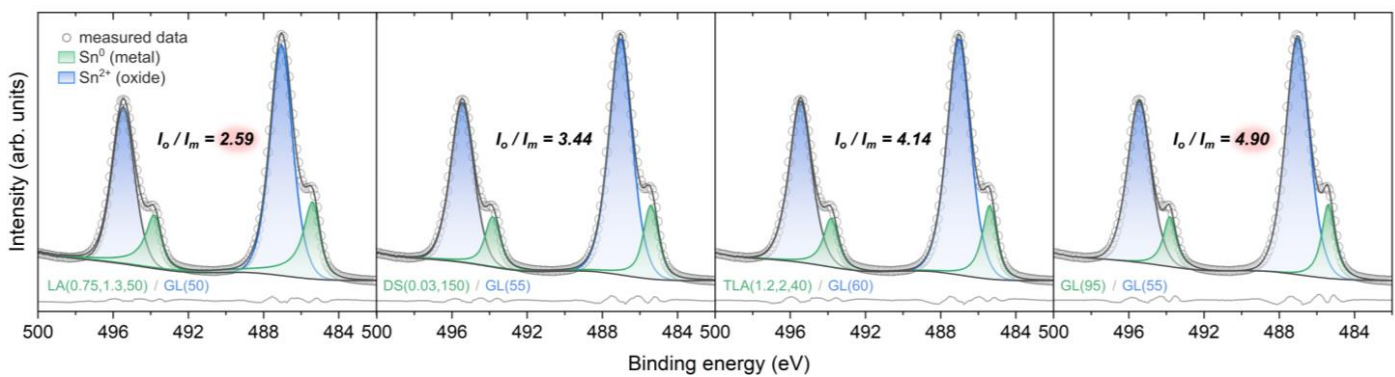

**Figure S1.** XPS core-level spectrum of Sn $3d_{3/2}$ and $3d_{5/2}$. The measured data is identical in all tiles but different line shapes and minimal background adjustments have been chosen to fit the data. GL(x) is the line shape of the $Sn^{2+}$ peak and denotes a convolution of a Lorentzian and a Gaussian, where $x$ is the percentage of Lorentzian contribution. For the metallic $Sn^0$ peak different line shapes were chosen: DS(a,b) is a Doniach-Sunjic profile, for which $a$ is an asymmetry parameter and $b$ defines the width of the Gaussian with which the function is convoluted. TLA(c,d,e) denotes an asymmetric Lorentzian with adjustable tail, with $c$ being an asymmetry parameter, $d$ dampens the asymmetric tail, and $e$ defines the width of the Gaussian with which the function is convoluted.

## 2. Transmission Electron Microscopy

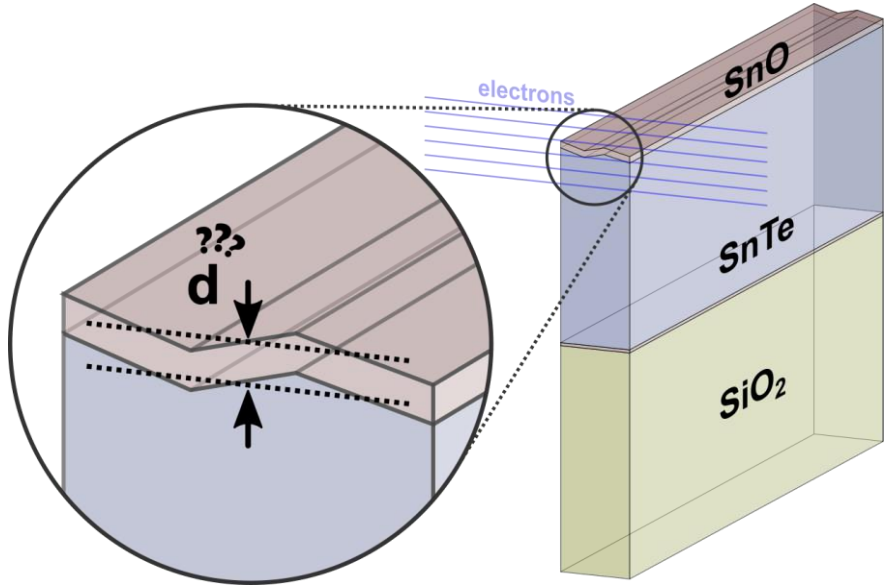

**Figure S2.** Drawing of the cross-section of a TEM-lamella, illustrating why the depth profile of an oxide overlayer of a rough thin film cannot be determined with high resolution and is likely to be overestimated.

## 3. X-ray Reflectivity

The fitting of the XRR data was performed in GenX[1] version 2.4.9 (homepage: http://genx.sf.net). The densities of SnO and SnTe are very similar, which is why the exact values obtained from the XRR model are unreliable. During fitting, the values varied more than the apparent deviation between the densities given in Table S1 and their corresponding literature values for bulk material ($\rho_{t-SnO} \approx 6.29$ g/cm$^3$ ≙ 1.63 e$^-$/Å$^3$ and $\rho_{c-SnTe} \approx 6.32$ g/cm$^3$ ≙ 1.59 e$^-$/Å$^3$ according to materialsproject.org[2]). It is not possible to draw definite conclusions from this model, because multiple plausible parametrizations are consistent with the measured data. The literature values for the densities of bulk materials were therefore used to calculate the XPS models.

**Table S1:** Fitting parameters of the XRR model shown in Figure 2 in the paper: thickness $d$, roughness $\sigma$, and density $\rho$ (GenX uses formula unit $fu$ per Å$^3$, proportional to electron density, which can be converted to mass density by $\rho \left[\frac{g}{cm^3}\right] = \rho_{fu} \left[\frac{fu}{Å^3}\right] \cdot M \left[\frac{g}{mol}\right] \cdot 1.66054$).

|  | SnO_overlayer | SnTe | SnO_substrate | SiO$_2$ |
| --- | --- | --- | --- | --- |
| $d$ (nm) | 2.50 | 48.93 | 0.30 | 47.65 |
| $\sigma$ (nm) | 0.42 | 1.25 | 0.10 | 0.45 |
| $\rho$ (g/cm$^3$) | 6.04 | 6.54 | 6.48 | 2.10 |

## 4. Mathematical Expressions

The following equations S1-S3 are the same as eq. 1-3 in the paper, shown here with the integrals of the depth distribution functions from which they were derived. The integrals are illustrated in Figure S3. Eq. 1/S1 is well known and widely used for native oxide overlayers:

$$\frac{I_{Sn^{2+}} \cdot N_{SnTe}}{I_{Sn^0} \cdot N_{SnO}}(L_{SnO}, L_{SnTe}, \theta) = \frac{\int_0^d e^{\frac{-z}{L_{SnO} \cdot \cos\theta}} dz}{\int_d^\infty e^{\frac{d-z}{L_{SnTe} \cdot \cos\theta}} e^{\frac{-d}{L_{SnO} \cdot \cos\theta}} dz} = \frac{L_{SnO}}{L_{SnTe}} \left( e^{\frac{d}{L_{SnO} \cdot \cos\theta}} - 1 \right) \quad (S1)$$

Here, $I_{Sn^{2+}}$ and $I_{Sn^0}$ are the measured signal contributions and $N_{SnO}$ and $N_{SnTe}$ are the atomic number densities of $Sn^{2+}$ in the oxide and $Sn^0$ in the metal, respectively. $L_{SnTe}$ and $L_{SnO}$ are the effective attenuation lengths (EAL) in the oxide and in the metal, $\theta$ is the emission angle, $d$ is the oxide layer thickness.

Eq. 2/S2 was first published in Ref. [3]:

$$\frac{I_{Sn^{2+}} \cdot N_{SnTe}}{I_{Sn^0} \cdot N_{SnO}}(L_{av}, \theta) = \frac{\int_0^\infty e^{\frac{-z}{L_{av} \cdot \cos\theta}} \operatorname{erfc}(k(z-d))\, dz}{\int_0^\infty e^{\frac{-z}{L_{av} \cdot \cos\theta}} \operatorname{erfc}(k(d-z))\, dz}$$

$$= \frac{\operatorname{erfc}(-kd) - \operatorname{erfc}\left(\frac{1}{2L_{av}k \cdot \cos\theta} - kd\right) \cdot \exp\left(\frac{1}{(2L_{av}k \cdot \cos\theta)^2} - \frac{d}{L_{av} \cdot \cos\theta}\right)}{\operatorname{erfc}(kd) + \operatorname{erfc}\left(\frac{1}{2L_{av}k \cdot \cos\theta} - kd\right) \cdot \exp\left(\frac{1}{(2L_{av}k \cdot \cos\theta)^2} - \frac{d}{L_{av} \cdot \cos\theta}\right)}. \tag{S2}$$

Here, the EALs $L_{SnTe}$ and $L_{SnO}$ were averaged ($L_{av}$) and the resulting depth distribution function $\varphi_{av}(z)$ was convoluted with an decreasing and an increasing error function for the oxide and the metal, respectively. Eq. 3/S3 is the revised version of eq. 2/S2, for which $L_{SnO}$ and $L_{SnTe}$ are considered as distanced quantities again:

$$\frac{I_{Sn^{2+}} \cdot N_{SnTe}}{I_{Sn^0} \cdot N_{SnO}}(L_{SnO}, L_{SnTe}, \theta) = \frac{\int_0^\infty e^{\frac{-z}{L_{SnO} \cdot \cos\theta}} \operatorname{erfc}(k(z-d))\, dz}{\int_0^\infty e^{\frac{d-z}{L_{SnTe} \cdot \cos\theta}} e^{\frac{-d}{L_{SnO} \cdot \cos\theta}} \operatorname{erfc}(k(d-z))\, dz}$$

$$= \frac{L_{SnO}}{L_{SnTe}} \cdot \exp\left(\frac{d}{L_{SnO} \cdot \cos\theta} - \frac{d}{L_{SnTe} \cdot \cos\theta}\right) \tag{S3}$$

$$\cdot \frac{\operatorname{erfc}(-kd) - \operatorname{erfc}\left(\frac{1}{2L_{SnO}k \cdot \cos\theta} - kd\right) \cdot \exp\left(\frac{1}{(2L_{SnO}k \cdot \cos\theta)^2} - \frac{d}{L_{SnO} \cdot \cos\theta}\right)}{\operatorname{erfc}(kd) + \operatorname{erfc}\left(\frac{1}{2L_{SnTe}k \cdot \cos\theta} - kd\right) \cdot \exp\left(\frac{1}{(2L_{SnTe}k \cdot \cos\theta)^2} - \frac{d}{L_{SnTe} \cdot \cos\theta}\right)}$$

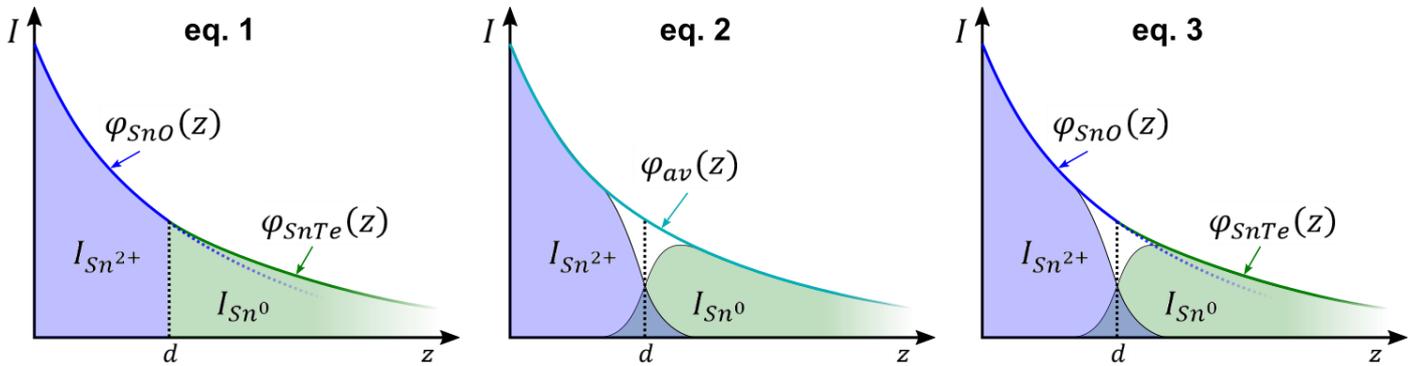

**Figure S3.** Schematic illustrations of the integrals of the depth distribution functions $\varphi(z)$, which were solved to derive eq. 1-3/S1-S3. $\varphi_{SnO}(z)$ is the integrand of the numerator and $\varphi_{SnTe}(z)$ the integrand of the denominator of the respective equation.

## 5. XPS Peak Fitting

Peak fitting of the XPS data was performed in CasaXPS version 2.3.22PR1.0. All metallic component peaks (denoted by superscript 0) have been fitted using different asymmetric peak functions (Table S2) as determined from the sputter-cleaned metallic reference. All other component peaks, such as oxides (denoted by superscript 2+) and satellite peaks, have been fitted using symmetric Voigt functions, which is a convolution of Gaussian and Lorentzian. The intensity ratios of all spin-orbit split peak doublets, both oxide and metal, were fixed to 1:2 for *p*-orbitals and 2:3 for *d*-orbitals.

Shirley or Tougaard background functions have been used to subtract the inelastic background signal. At higher binding energies the backgrounds had to be modified by spline functions to account for the nonlinear background signal at the beginning and end of the main emission peaks. All binding energy positions, line shapes and background types are given in Table S2. The unweighted intensity ratios $I_{Sn^{2+}}/I_{Sn^0}$ (i.e. without atomic number densities) as used in CasaXPS are shown in Table S3.

**Table S2.** Binding energy positions (with small variations of about ±0.1 eV), line shapes and background types used in the fitting of all emission peaks. The line shape type GL denotes a convolution of a Gaussian and a Lorentzian peak. LA denotes a Lorentzian asymmetric line shape.

| Orbital | Binding Energy (eV) | Line shape | Background type |
|---|---|---|---|
| $O^{2-}$ 2s | 21.9 | GL(30) | |
| O 2s (O–C) | 23.5 | GL(30) | |
| $Sn^0$ $4d_{5/2}$ | 24.4 | LA(1.25,2.5,25) | Shirley |
| $Sn^0$ $4d_{3/2}$ | 25.3 | LA(1.25,2.5,25) | |
| $Sn^{2+}$ $4d_{5/2}$ | 26.3 | GL(30) | |
| $Sn^{2+}$ $4d_{3/2}$ | 27.4 | GL(30) | |
| $Te^0$ $4d_{5/2}$ | 39.8 | LA(0.9,1.5,150) | Spline Shirley |
| $Te^0$ $4d_{3/2}$ | 41.3 | LA(0.9,1.5,150) | |
| $Sn^0$ 4s | 137.9 | LA(1,3,150) | Shirley |
| $Sn^{2+}$ 4s | 138.4 | GL(50) | |
| primary C 1s | 284.9 | - | - |
| $Sn^0$ $3d_{5/2}$ | 485.3 | LA(1,1.3,50) | Spline Shirley |
| $Sn^{2+}$ $3d_{5/2}$ | 486.9 | GL(55) | |
| $Sn^0$ $3d_{3/2}$ | 493.7 | LA(1,1.3,50) | |
| $Sn^{2+}$ $3d_{3/2}$ | 495.4 | GL(55) | |
| $O^{2-}$ 1s | 530.6 | GL(30) | Linear |
| O 1s (O–C) | 532.0 | GL(40) | |
| $Te^0$ $3d_{5/2}$ | 572.4 | LF(1.5,2.5,8,50) | Spline Shirley |
| $Te^0$ $3d_{3/2}$ | 582.8 | LF(1.5,2.5,8,50) | Spline Tougaard |
| $Sn^0$ $3p_{3/2}$ | 715.3 | LA(0.9,1.3,150) | 2x Spline Shirley |
| $Sn^{2+}$ $3p_{3/2}$ | 716.7 | GL(90) | |
| $Sn^0$ $3p_{3/2}$ (sat) | 733.0 | GL(30) | |
| $Sn^0$ $3p_{1/2}$ | 756.8 | LA(0.9,1.3,150) | |
| $Sn^{2+}$ $3p_{1/2}$ | 758.6 | GL(90) | |
| $Sn^0$ $3p_{1/2}$ (sat) | 774.3 | GL(30) | |
| $Te^0$ $3p_{3/2}$ | 819.1 | LA(0.9,1.5,100) | 2x Spline Shirley |
| $Te^0$ $3p_{3/2}$ (sat) | 836.3 | GL(30) | |
| $Te^0$ $3p_{1/2}$ | 869.9 | LA(0.9,1.5,100) | |
| $Te^0$ $3p_{1/2}$ (sat) | 885.4 | GL(30) | |
| $Sn^0$ 3s | 883.4 | LA(1,3,150) | |
| $Sn^{2+}$ 3s | 886.3 | GL(80) | |

**Table S3:** Unweighted intensity ratios $I_{Sn^{2+}}/I_{Sn^0}$ for all angles and peak regions.

| θ | 4d | 4s | 3d | 3p | 3s |
|---|---|---|---|---|---|
| 75 | 0.066 | 0.060 | 0.040 | 0.026 | 0.018 |
| 60 | 0.177 | 0.161 | 0.110 | 0.077 | 0.057 |
| 45 | 0.289 | 0.263 | 0.185 | 0.132 | 0.100 |
| 30 | 0.379 | 0.347 | 0.246 | 0.179 | 0.137 |
| 15 | 0.488 | 0.401 | 0.287 | 0.210 | 0.161 |
| 0 | 0.458 | 0.419 | 0.301 | 0.221 | 0.170 |

## 6. Calculation of Material Properties

There is a consensus that the effective attenuation length (EAL) $L$ is better suited for quantitative XPS analyses than the inelastic mean free path (IMFP) $\lambda$, because it also takes the inelastic scattering into account. Therefore, only EALs were used in the paper. The calculation of both variables is shown below for comparison. The equations are plotted in Figure S4.

The IMFPs, which were not used in the paper, can be calculated by the predictive TPP-2M formula, proposed by Tanuma, Powell, and Penn:[4]

$$\lambda = \frac{E_k}{E_p^2 \left[\beta \ln(\gamma E_k) - \ln(C/E_k) + (D/E_k^2)\right]} \tag{S4}$$

$$\beta = -0.1 + 0.944 \sqrt{E_p^2 + E_g^2} + 0.069 \rho^{0.1} \tag{S5}$$

$$\gamma = 0.191 \sqrt{\rho} \tag{S6}$$

$$C = 1.97 - 0.91 U \tag{S7}$$

$$D = 53.4 - 20.8 U \tag{S8}$$

$$U = \frac{N_v \rho}{M} = E_p^2 / 829.4 \tag{S9}$$

Here, $E_k$ is the kinetic energy of the photoelectron in eV, $E_p$ is the free electron plasmon energy in eV, $N_v$ is the number of valence electrons per atom or molecule, $\rho$ is the density in g cm$^{-3}$, $M$ is the atomic or molecular weight and $E_g$ is the band gap in eV. Table S4 shows the values used to calculate the IMFPs using eq. S4-S9 for SnTe and SnO.

Just for comparison, we also applied the so-called universal curve for IMFP in inorganic compounds, as proposed by Seah and Dench:[5]

$$\lambda = \frac{641}{E_k^2} + 0.096 \cdot \sqrt{E_k} \tag{S10}$$

The EALs were calculated by a universal curve proposed by Seah:[6]

$$L = \frac{(5.8 + 0.0041 \cdot Z^{1.7} + 0.088 \cdot E_k^{0.93}) \, a^{1.82}}{Z^{0.38}(1 - 0.02 \cdot E_g)}, \tag{S11}$$

with the average atomic number $Z = \frac{g Z_g + h Z_h}{g+h}$ of a binary compound G$_g$H$_h$ with stoichiometry coefficients $g$ and $h$ and corresponding atomic numbers $Z_g$ and $Z_h$. $E_g$ is the band gap in eV, $E_k$ the kinetic energy of the photoelectron in eV and $a$ the thickness per monolayer in nm:[6]

$$a = \sqrt[3]{\frac{M}{\rho \, N_A \, (g+h)}} \tag{S12}$$

Thereby, $M$ is the molar mass, $\rho$ the mass density and $N_A$ the Avogadro constant. The material parameters that were used in the calculation of the IMFPs and EALs are given in Table S4.

**Table S4:** Material parameters for fitting the XPS data. The atomic number density $N = \rho \, N_A / M$ was used for weighting of the intensity ratios.

|  | $M$ (u) | $\rho$ (g cm$^{-3}$) | $N_v$ | $E_g$ (eV) | $N$ (10$^{22}$ cm$^{-3}$) | $Z$ | $a$ (nm) |
|---|---|---|---|---|---|---|---|
| SnTe | 246.31 | 6.32 | 10 | 0.18 | 1.55 | 51 | 0.32 |
| SnO | 134.71 | 6.29 | 10 | 0.7 | 2.81 | 29 | 0.26 |

From these values we calculated the IMFPs and EALs as functions of $E_{kin}$ for SnTe and SnO (Figure S4). For spin-orbit split peaks, the average $E_{kin}$ was used, weighted by the respective intensity ratio (1:2 for *p*-orbitals and 2:3 for *d*-orbitals). Table S5 shows the results.

**Table S5:** Calculated values of IMFPs and EALs of SnTe and SnO for all examined peak regions.

| Peak assignment | $\lambda_{SnO}$ (Å) | $\lambda_{SnTe}$ (Å) | $L_{SnO}$ (Å) | $L_{SnTe}$ (Å) |
|---|---|---|---|---|
| 4d | 26.1 | 30.8 | 20.61 | 24.26 |
| 4s | 24.5 | 28.9 | 19.25 | 22.70 |
| 3d | 19.5 | 22.9 | 14.96 | 17.75 |
| 3p | 15.8 | 18.6 | 11.95 | 14.34 |
| 3s | 13.4 | 15.8 | 9.98 | 12.07 |

There has been an extensive debate regarding the accuracy of different experimental and theoretical approaches used to determine the IMFP and EAL. While various methods lead to some degree of quantitative uncertainty, experimental and theoretical results have generally shown reasonable agreement.[7] The calculations presented here do not aim to contribute to this nuanced debate or challenge findings from more detailed studies that may already exist.

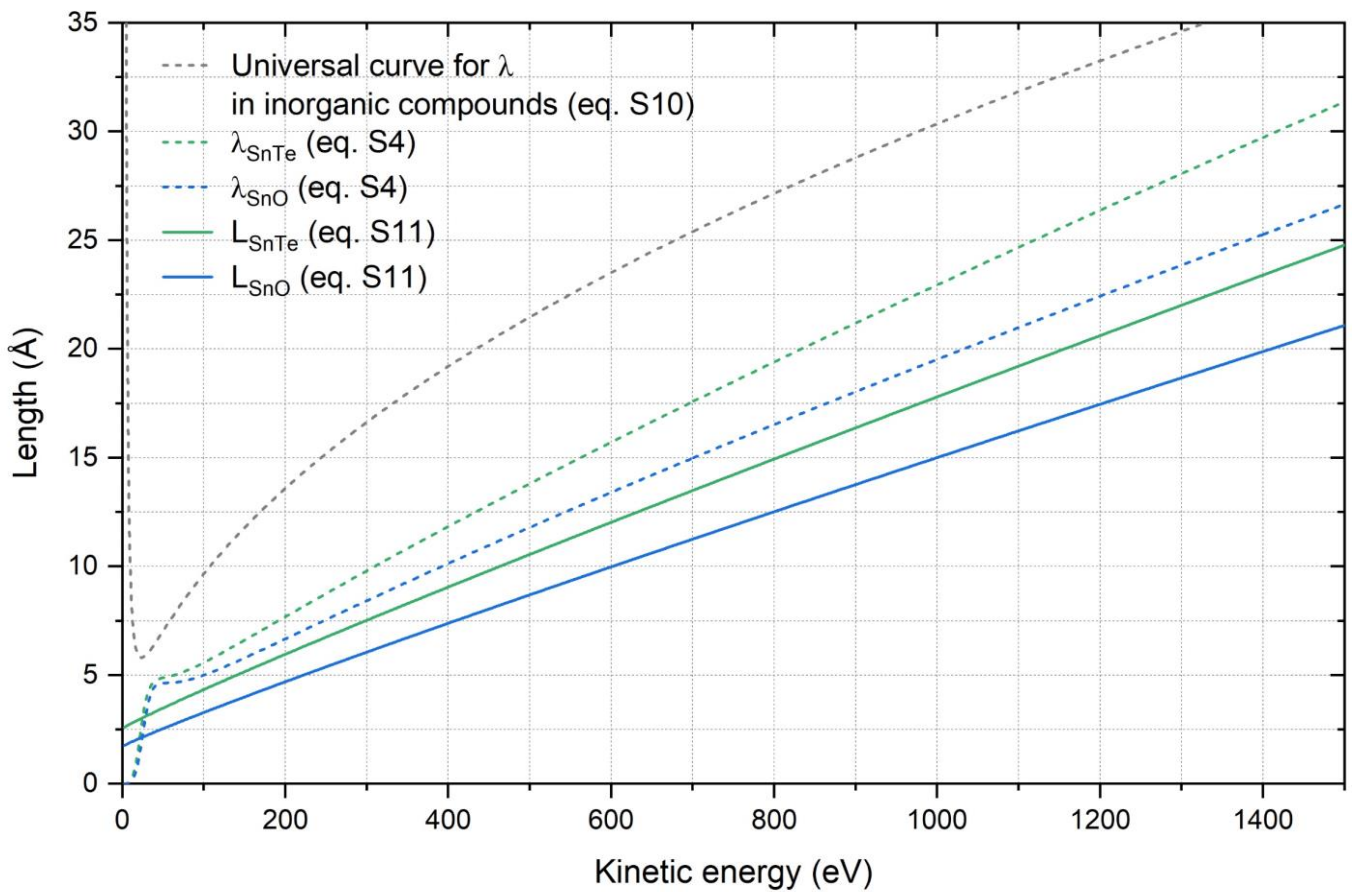

**Figure S4.** IMFP $\lambda(E_k)$ (dashed lines) and EAL $L(E_k)$ (solid lines).

## 7. Auger Peaks by AES and XPS

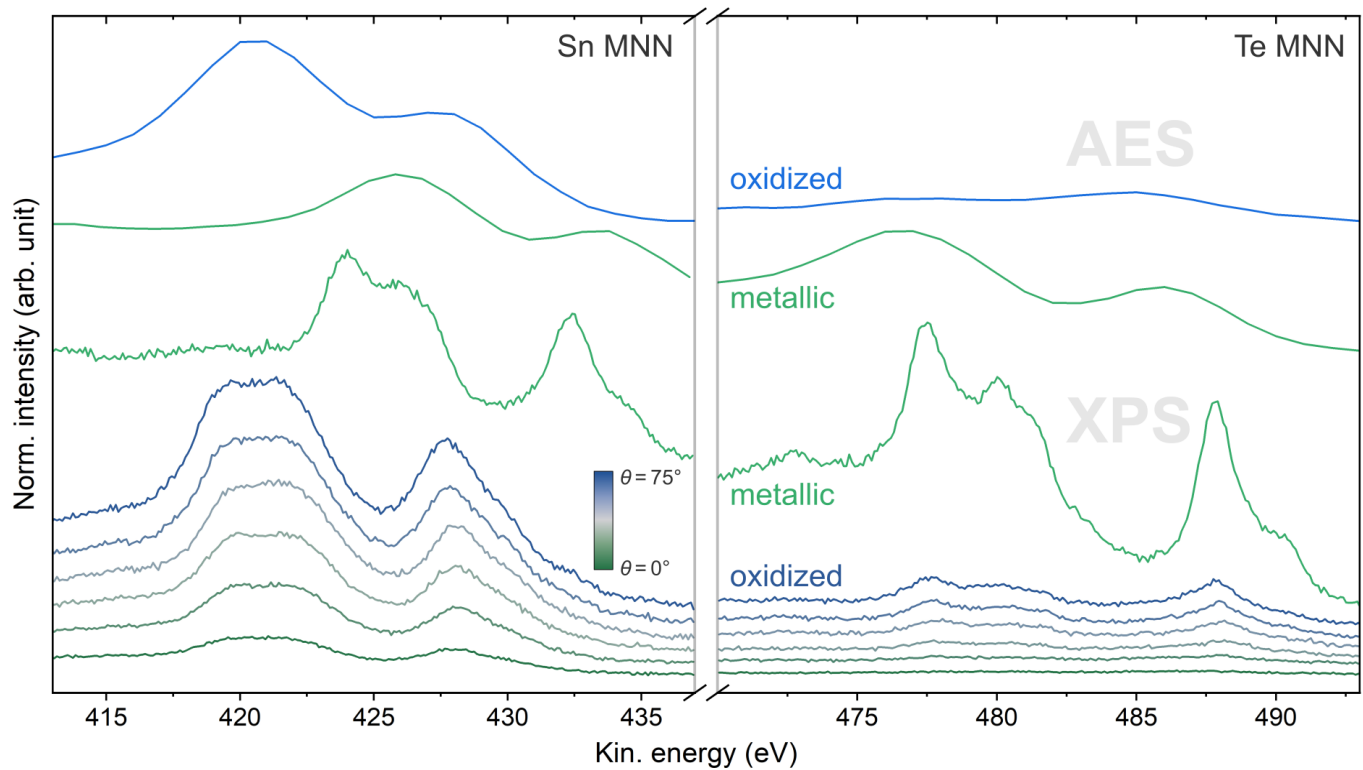

**Figure S5.** Sn MNN and Te MNN peaks measured by (AR-)XPS (7 curves at bottom) and AES (2 curves at top) before sputtering ("oxidized") and after sputtering ("metallic"). The relative scaling was adjusted for clarity. The (AR-)XPS spectra have the same scaling (not normalized). All XPS spectra have been referenced to the Fermi level on the binding energy scale and were then converted back to the kinetic energy scale using the following equation: $E_k = h\nu - E_b - \phi$.

## 8. Influence of the Interface Sharpness

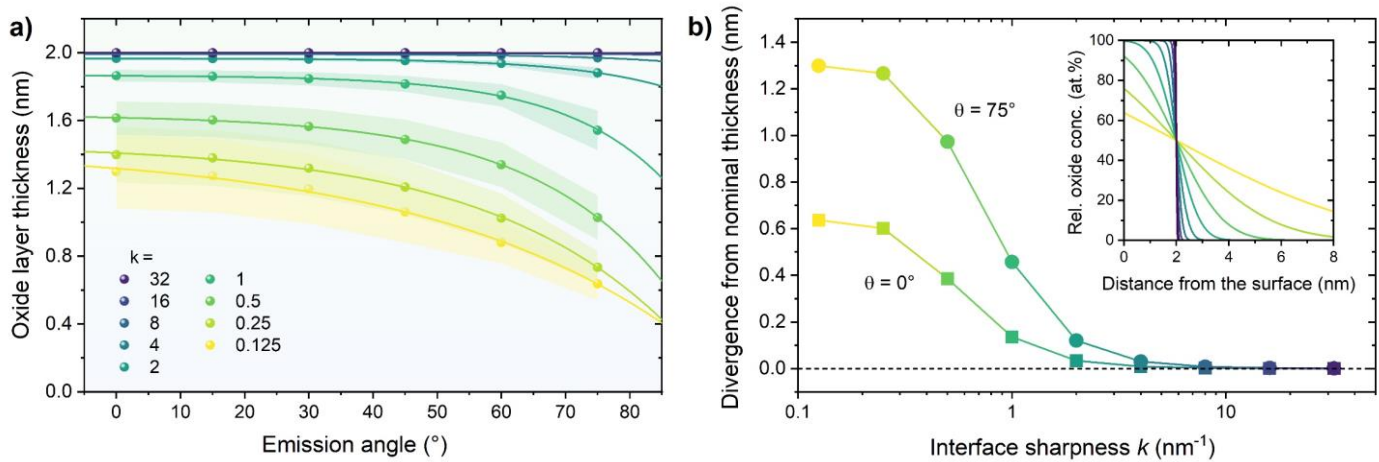

**Figure S6.** Systematic underestimation of the overlayer thickness by the single-energy model with increasing interface broadness: **(a)** Theoretical SnO overlayer thickness on SnTe according to the single-energy model (eq. 1) as a function of the emission angle. The intensity ratios on which these results are based were calculated using the multiple-energies model (eq. 3) with constant $d = 2$ and varying $k$ values in nm$^{-1}$. Shaded regions indicate standard deviation of the results from different emission peaks. The lines are fits ($y = a - bc^x$) and serve as a guide to the eye. **(b)** Difference between the result of the single-energy model (eq. 1) and the nominal oxide layer thickness $d = 2$ as a function of interface sharpness $k$. The color scheme is the same as in (a). The inset shows the corresponding concentration profiles defined by $C_{Sn^{2+}}(z) = 0.5 \cdot erfc(k \cdot (z - d)) \cdot 100\%$. Mind that profiles with values of $k < 0.5$ should not be modelled as two distinct layers with defined EAL and atomic number densities. Since eq. 3 reduces to eq. 1 in the limit $k \to \infty$, it is not surprising that eq. 1 yields increasingly accurate results for increasing values of $k$, i.e. sharper interfaces.

# References

[1] Glavic, A., & Björck, M. (2022). GenX 3: the latest generation of an established tool. Applied Crystallography, 55(4), 1063-1071.

[2] Jain, A., Ong, S. P., Hautier, G., Chen, W., Richards, W. D., Dacek, S., ... & Persson, K. A. (2013). Commentary: The Materials Project: A materials genome approach to accelerating materials innovation. APL materials, 1(1).

[3] Wortmann, M., Viertel, K., Westphal, M., Graulich, D., Yang, Y., Gärner, M., ... & Kuschel, T. (2023). Sub-Nanometer Depth Profiling of Native Metal Oxide Layers Within Single Fixed-Angle X-Ray Photoelectron Spectra. Small Methods, 2300944.

[4] Tanuma, S., Powell, C. J., & Penn, D. R. (2003). Calculation of electron inelastic mean free paths (IMFPs) VII. Reliability of the TPP-2M IMFP predictive equation. Surface and interface analysis, 35(3), 268-275.

[5] Seah, M. P., & Dench, W. A. (1979). Quantitative electron spectroscopy of surfaces: A standard data base for electron inelastic mean free paths in solids. Surface and interface analysis, 1(1), 2-11.

[6] Seah, M. P. (2012). Simple universal curve for the energy-dependent electron attenuation length for all materials. Surface and interface analysis, 44(10), 1353-1359.

[7] Jablonski, A., & Powell, C. J. (2020). Effective attenuation lengths for different quantitative applications of X-ray photoelectron spectroscopy. Journal of Physical and Chemical Reference Data, 49(3).